\documentclass[aps,prl,reprint,twocolumn,showpacs,unsortedaddress]{revtex4-1}

\usepackage{amsmath,amssymb}
\usepackage{pstricks}
\usepackage{mathrsfs}

\usepackage[dvips]{graphicx}
\usepackage[Symbol]{upgreek}

\usepackage{bm}
\usepackage{color}

\begin{document}

\title{Vortex-state-mediated Josephson effect}

\begin{abstract}

 The Josephson effect is a kind of macroscopic quantum phenomenon that supercurrent flows through a Josephson junction without any voltage applied. We predict a novel vortex-state-mediated Josephson effect in an SNS  Josephson junction supporting vortices.  The vortex-state-mediated supercurrent is enhanced or reduced significantly in magnitude depending on the junction length, and exhibits several steps with the number of effective propagating channels in current-phase evolution at zero temperature.  At finite temperatures, these supercurrent steps persist in the short junction limit, and develop into sawtooth oscillations if the junction length becomes comparable to the coherence length $\xi=\hbar v_F/\Delta$ of the superconductor, and in later case a supercurrent reversal can be observed.  These findings may provide a smoking-gun signature of vortex bound states in superconductors and promise possible applications in future Josephson devices.
\end{abstract}

\pacs{74.50.+r,74.45.+c, 73.20.-r,73.40.Gk}

\author{Zhao Yang Zeng}
\email{zyzeng@jxnu.edu.cn}
\author{Chong Zhang}
\affiliation{Department of Physics, Jiangxi Normal  University, Nanchang 330022, China}

\date{\today}

\maketitle

The Josephson effect is a kind of macroscopic quantum phenomenon, first predicted by Brian Josephson that the cooper pairs can tunnel through weakly coupled superconductors\cite{Josephson}. It also exists if superconductors are connected by a weak link of any physical nature\cite{Gennes,Likharev}.

In a clean SNS junction, the Josephson effect is remarkably different from the Josephson tunnel junctions. The underlying mechanism is the coherent Andreev refletion\cite{Andreev}. Interference between  Andreev reflected electronlike and holelike excitation wavefunctions in the quantum well formed by the pair potentials of the superconductors leads to the formation of  Andreev bound states\cite{Kulik}. It is shown that a significant portion of  supercurrent is carried by the discrete Andreev levels\cite{Bagwell,Tang}, and the critical current(the maximum current) decreases exponentially with temperature\cite{Kulik}. Introduction of impurities in the normal region would suppress the Josephson supercurrent\cite{Bagwell}. If the normal region becomes a quantum point contact\cite{superconductor-point}, the critical current may be an integer multiple of $e\Delta/\hbar$, where $2\Delta$ is the energy gap of the superconductor.

The existence of bound states inside a vortex, where the pair potential of the superconductor is zero, is predicted by Caroli, de Gennes, and Matricon\cite{Caroli} and confirmed with controversy\cite{Hess} long after its discovery. The vortex bound states are still  the Andreev bound states. Vortices in chiral P-wave superconductors or superfluid may even support zero-energy majorana modes\cite{Majorana,Fu1},  one of the appealing candidates for topological quantum computation\cite{Nayak}. Tunneling between the majorana zero modes at two vortices are investigated\cite{Majorana tunneling}, demonstrating that the tunneling amplitude depends on the phase difference of the order parameters at the two vortices and decays exponentially with the distance between the vortices.

A question arises that what happens when both the vortex bound states and Andreev bound states appear in an SNS junction. One may suggest intuitively that the tunneling between vortex bound states also carry a portion of supercurrent, besides the Andreev bound states, at least in the short junction case. So an enhancement of the supercurrent in the presence of vortex bound states can be expected.  For this purpose, we propose an SNS junction that can support both the vortex bound states and the Andreev bound states, and investigate the supercurrent-phase characteristics in such a particular setup. The proposed SNS junction is very similar to the hybrid structures based on topological insulators\cite{Ran,Ioselevich}. It is known that the surface of a topological insulator can be described by a Dirac Hamiltonian\cite{Topological}. In this sense, our proposed SNS Josephson junction can be considered as the $Schr\ddot{o}$dinger version of  the similar topological SNS junctions\cite{Ran, Ioselevich}.

\begin{figure}
\centering
		\includegraphics[width=0.4\textwidth]{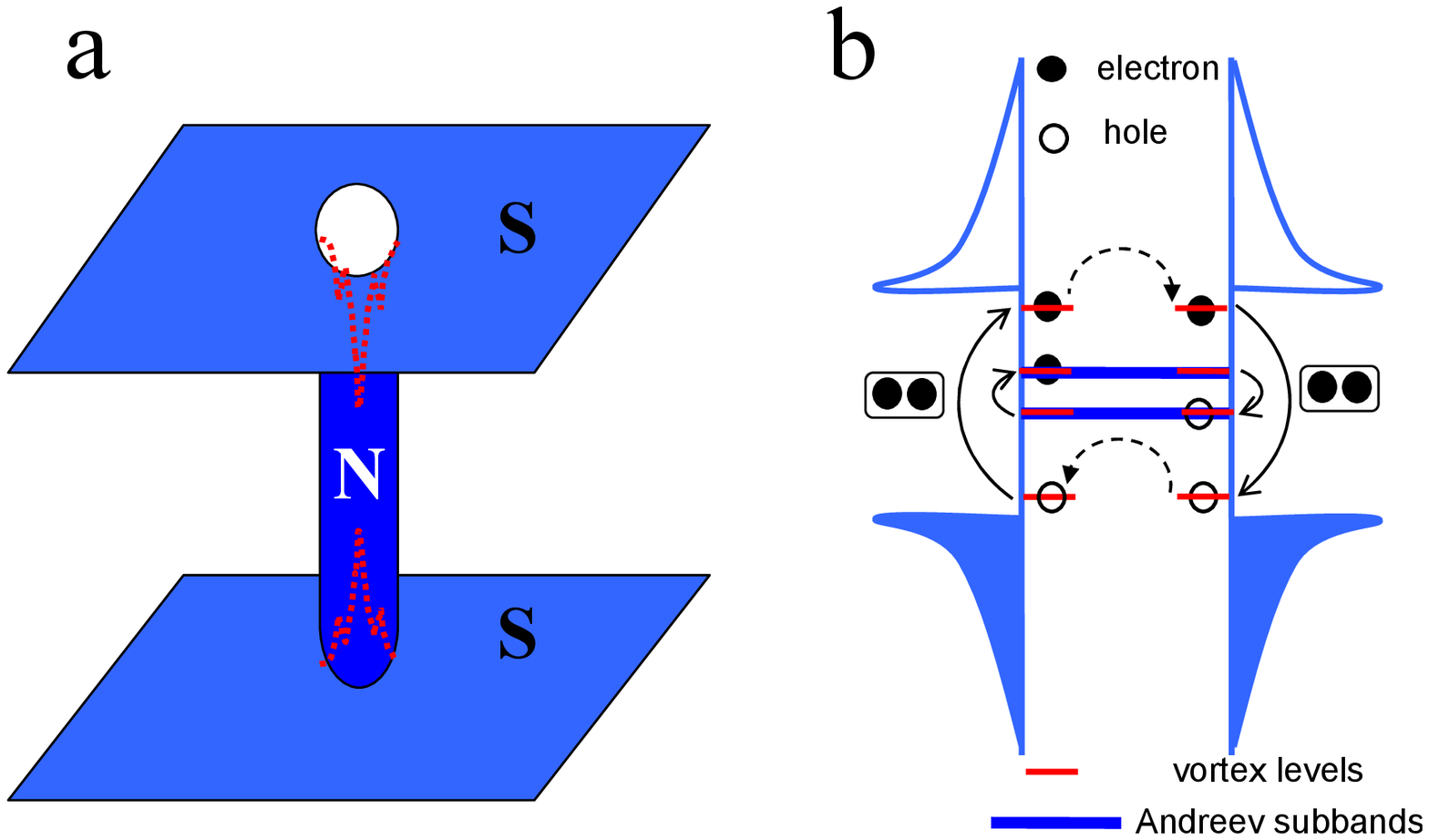}
  \caption{(a) SNS junction with vortices. Two superconductive planar slabs, each with a hole of radius $R$, are connected ideally to a normal cylindrical slab of radius $R$ and length $L$ to form an SNS junction. The red peaked lines are sketched to represent the vortex bound states localized inside the hole region. (b) Schematic view of transport processes of cooper pairs. Cooper pairs can be transported from one superconductive slab to the other through the  Andreev subbands (blue lines) on the cylindrical surface, or by coherent tunneling via vortex bound levels (red lines). Note that  the vortex levels can be hybridized with the Andreev subbands if their angular momenta are coincident.}
\label{model}
\end{figure}

The SNS Josephson junction under consideration consists of two s-wave superconductive planar slabs, each with a hole, and  a hollow cylindrical normal slab. The superconductive slabs are connected ideally by the normal slab(see Fig. $\ref{model}$(a)). Here "ideally" means that neither barriers at the junctions nor  Fermi velocity mismatch between the superconductor and the normal conductor are considered.  Unlike the Dirac version of the junction based on topological insulators with insulating bulk  and conducting surface, it may pose an experimental challenge to realize the proposed SNS junction. One possibility of realization is to prepare a shaped two-dimensional electron gas($2$DEG) by  heterostructure($GaAs/Al_{x}Ga_{1-x}As$) engineering, and then introduce superconductivity at the $2$DEG planes by the proximity effect\cite{Fu1}.

The surface  superconductor in the presence of a vortex with a flux quantum $\Phi_0=hc/2e$ can be described by the Bogoliubov-de Gennes Hamiltonian\cite{Gennes} with an inhomogeneous pair potential $\Delta(\bm{\rho})$

\begin{equation}
H_{S}=\left[
\begin{array}{cccc}
\mathscr{H}_e
&\Delta(\bm{\rho}) \\
\Delta^*(\bm{\rho}) &
\mathscr{H}_h
\end{array}
\right],
\label{BdG-Hamiltonian}
\end{equation}
where $ \mathscr{H}_e =-\hbar^2(\mathbf{e}_\rho \partial_\rho +\mathbf{e}_\theta \partial_\theta /\rho+i
e A_\theta \mathbf{e}_\theta  /\hbar c)^2/2m-E_F$, $ \mathscr{H}_h=-\mathscr{H}_e^*$ are the single-electron and single-hole Hamiltonian.  We choose a gauge such that the pair potential takes the form $\Delta(\bm{\rho})=\Delta(\rho)e^{-i\theta}$, with $\Delta(\rho)\propto \rho$ as $\rho\rightarrow 0 $ and $\Delta(\rho)= \Delta $ at distances $\rho>\xi$.  Expressing the  two-component wavefunctions in the  form $\Psi_S(\rho,\theta)=\exp(
i\mu\theta-i\theta\sigma _z/2)[u(\rho),v(\rho)]^T$, we obtain the following Bogoliubov-de-Gennes equation

\begin{eqnarray}
\Big\{-\frac{\hbar^2}{2m}
\Big[&&\frac{1}{\rho}\frac{d}{d\rho}\Big(\rho\frac{d}{d\rho}\Big)-\frac{1}{\rho^2}\Big(\mu-\frac{\sigma_z}{2}+
\frac{\sigma_ze\rho}{\hbar c}
A_\theta(\rho)\Big)^2+k_F^2\Big]  \nonumber \\
 &&\hspace{2cm}\sigma_z+ \Delta(\rho)\sigma_x -E
  \Big\} \Big[\begin{array}{l} u(\rho)\\v(\rho)\end{array}\Big]=0,
\label{BdG-equation}
\end{eqnarray}
where $ e\rho A_\theta(\rho)/\hbar c = H \rho^2 /2\Phi_0 \sim H \xi^2/2\Phi_0 \sim H/H_{c2}\ll1$ and therefore the magnetic field effect can be safely neglected. Unfortunately Eq. (\ref{BdG-equation}) can not be solved exactly, we resort to approximate solutions following the perturbation treatment of Calori et al.\cite{Caroli} and find\cite{Supplement material}

\begin{eqnarray}
\Psi_S(\rho,\theta)=\sum_\mu && e^{i(\mu\theta-i\sigma_z\theta/2)}\Big\{C_\mu^1H^{1}_{\mu'}(k_F\rho)\Big[\begin{array}{l} e^{i\Re(\rho)} \\e^{-i\Re(\rho)}\end{array}\Big]\nonumber\\
&&\hspace{0.5cm}+C_\mu^2H^{2}_{\mu'}(k_F\rho)\Big[\begin{array}{l} e^{-i\Re(\rho)} \\e^{i\Re(\rho)}\end{array}\Big]\Big\}e^{-\Im(\rho)},
\label{S-function}
\end{eqnarray}
with $H^{1,2}_{\mu'}$ being Hankel functions of the first and second kind indexed by $\mu'=(\mu^2+1/4)^{1/2}$  and $\Im(\rho)=\int_0^\rho \Delta(r)dr/\hbar v_F$. Single valuedness of wavefunctions restricts  $\mu$ to take only half-integral values.  The term $\Re(\rho)=\pi/4+\Re_1=\pi/4-\int_{\rho}^{\infty} e^{2[\Im(\rho)-\Im(r)]}(E/ \hbar v_F+\mu/2k_Fr^2)dr$ represents a phase shift in the presence of the vortex.

Particles propagating on the normal cylindrical surface satisfy the Schr$\ddot{o}$dinger equation on the curved space\cite{Supplement material}
\begin{equation}
\sigma_z[-\frac{\hbar^2}{2m}(\partial_z^2+\frac{1}{R^2} \partial_\theta^2)-E_F]\Psi_C(\theta,z)=E\Psi_C(\theta,z).
\label{cylindrical equation}
\end{equation}
 The two-component wavefunctions are

\begin{equation}
\label{cylindrical wave function}
\Psi_C(z,\theta)=\sum_{\nu}e^{i\nu\theta}\Big[\begin{array}{l}  D^{1+}_{\nu}e^{ik_{+}^{\nu}z}+D^{1-}_{\nu}e^{-ik_{+}^{\nu}z} \\ D^{2+}_{\nu}e^{ik_{-}^{\nu}z}+D^{2-}_{\nu}e^{-ik_{-}^{\nu}z} \end{array}\Big],
\end{equation}
where $k_{\pm}^{\nu}=[2m(E_F\pm E)/\hbar^2-\nu^2/\rho^2]^{1/2}$ are the wave vectors of the electronlike and holelike excitations, respectively. Note that we have neglected the small magnetic field effect as well, since the magnetic field is  extended over a radius much greater than the cylinder radius $R$.

\begin{figure}
\centering
		\includegraphics[width=0.4\textwidth]{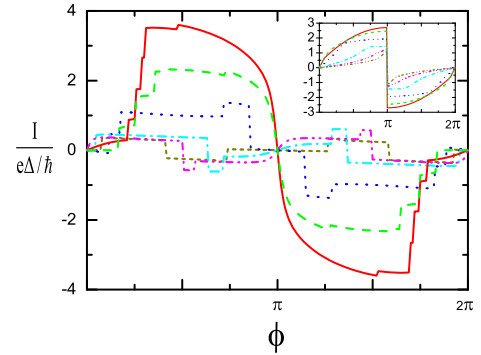}

\caption{ The supercurrent-phase relation $I(\phi)$ at zero temperature as $k_FR=2$ is computed for $k_FL = 1$ (red solid),  $k_FL = 2$ (green dashed), $k_FL = 5$ (blue dotted), $k_FL = 10$ (cyan dash-dotted),  $k_FL = 13$ (magenta dash-dot-dotted) and $k_FL = 15$ (dark yellow short-dashed).  Other parameters are chosen as $ k_F\xi=10, k_F\xi_{sc}=4.$  The results  of the gedenken junction with the same parameters are plotted in the inset.}
\label{current-phase-kR=2}
\end{figure}

  Excitation spectrum can be obtained by matching the wavefunctions on the upper and lower superconductive surfaces with the wavefunction on the normal cylindrical surface at the boundaries $\rho=R$. A detailed matching procedure is given in\cite{Supplement material}. The excitation energy  as a function of the superconducting phase difference $\phi$  with a fixed angular momentum $\mu=n\pm 1/2$  is approximately determined by the equation $(k_{+}^{\mu-1/2}-k_{-}^{\mu+1/2})L-4\Re_1(R)\pm\phi=(2n+1)\pi$.  Compared with the Andreev spectrum equation given by Eq. (\ref{gedenke-Andreev-spectrum}) in \cite{Supplement material} for a gedenken junction, one observes that the vortex bound state effects are encoded in an additional phase term $-4\Re_1(R)$. To estimate the value of the function $\Re_1(R)$, we approximate the pair potential near the vortex center by
$\Delta(\rho)\approx \Delta \rho/(\rho^2+2\xi^{2}_{sc})^{1/2}$, where $\xi_{sc}=(D/2\Delta)^{1/2}\ll\xi$ is the "dirty-limit" coherence length of the superconductive surfaces with diffusion constant $D$. In the regime of interest, $k_F R\gg1, R\ll\xi$, $\Re_1(R)\approx\Re_1(0)\approx -E /2\Delta+\mu\ln(\xi/\xi_{sc})/ k_F\xi$. The energy separation of vortex states is then estimated to be $\Delta^2/E_F$, consistent with the results of Calori et al..\cite{Caroli}

The Josephson current is an equilibrium property of superconductors, and can be expressed by\cite{Bardeen Beenakker,Supplement material}

\begin{equation}
I=-\frac{2e}{\hbar}\sum_{\mu}^{0< E_\mu <\Delta}tanh\Big(\frac{E_\mu}{2k_BT}\Big)\frac{\partial E_\mu}{\partial\phi},
\label{Josephson current}
\end{equation}
where the summation is over all the discrete subgap Andreev levels. At zero temperature, the Josephson supercurrent is
$I=-\frac{2e}{\hbar}\sum_{\mu;0< E_\mu}\frac{\partial E_\mu}{\partial\phi}$.

 Consider first the SNS junction without the cylinder connection. Two vortices can still host  bound states. The wavefunctions of the vortex bound states may extend somewhere along the axial direction of the cylinder, since any physical slabs have a finite thickness.  The bound state wavefunctions inside the vortex cores  are given by  $ \psi_{v}(\mathbf{r})\propto [ e^{i\frac{\phi}{2}}e^{i(\mu-\frac 12)\theta}J_{\mu-1/2},e^{-i\frac{\phi}{2}}e^{i(\mu+\frac 12)\theta}J_{\mu+1/2}]^T  \Phi_\mu(\rho)e^{-k_{F}^{\perp}|z|}$,
where $J_{\mu\mp1/2}$ are Bessel functions with argument $(k_\perp\pm m\epsilon_\mu/h^2k_\perp) \rho$, $\Phi_\mu$ is a decaying  function, $k_{F}^{\perp}\approx k^F[\mu^2/(k_FR)^2-1)]^{1/2}$. The amplitude for tunneling between the two vortices can be estimated from the Bardeen's well-know transfer hamiltonian method\cite{Bardeen} and is proportional to $\int d\rho d\theta \rho[\psi_{v}^{1\dagger} \partial \psi_{v}^{2}/\partial z- \partial \psi_{v}^{1\dagger} /\partial z \psi_{v}^{2}] \propto sin(\phi/2)e^{-k_F^\perp L}$.  From this expression we see that only the  vortex bound states with small angular momentum favor an effective tunneling. Adiabatic connection of the conductive cylindrical surface with the superconductive surfaces  opens up conductive channels to transport  cooper pairs by the formation of the Andreev subbands.  On the other hand, it leads to hybridization between the vortex bound states and the Andreev subbands on the cylindrical surface. The hybridization becomes most pronounced when the corresponding angular momenta are coincident.

\begin{figure}
\centering
		\includegraphics[width=0.4\textwidth]{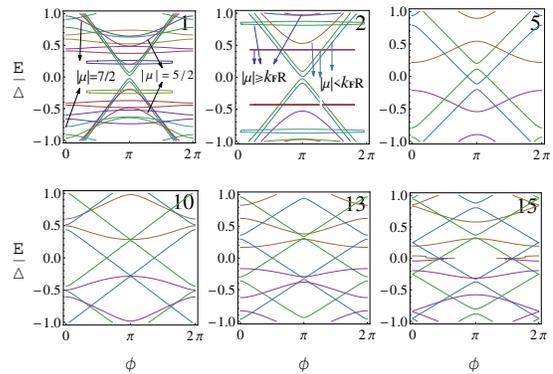}

\caption{Excitation spectra of the junction with different cylinder lengths $k_FL$ indicated by the number in the figures. The other parameters are the same  as in Fig. \ref{current-phase-kR=2}.}
\label{energy-spectra-KR=2}
\end{figure}

The supercurrent of a Josephson junction exhibits usual sinusoidal relation as a function of the superconducting phase difference. In Fig. $\ref{current-phase-kR=2}$ we present some examples of supercurrent-phase relation for the junction with fixed radius $k_FR=2$ and  different lengths $L$.  The evolution period of  supercurrent with the phase difference is still $2\pi$ as in the usual SNS junction case. The $2\pi$-periodicity of the supercurrent evolution is originated from $2\pi$-periodicity of the energy spectra.  At $\phi=\pm \pi$ the positive and negative Andreev modes have combined to form a standing wave with $\partial E^{\pm}/\partial \phi$\cite{Bagwell}, which are clearly shown in the excitation energy spectra in  Fig. \ref{energy-spectra-KR=2}. Therefore  the Josephson current drops to zero  at $\phi=\pm \pi$.

A striking phenomenon is observed that the supercurrent develops step structures in response to the increase of the superconducting phase difference. The number of steps within one half period is the same as the number of effective supercurrent-carrying modes, which can be confirmed from the excitation spectra in Fig. \ref{energy-spectra-KR=2}.     To the best of our knowledge, This observation has never been reported before. It can be considered as the fingerprint of the proposed junction.   By comparing the supercurrent in the presence and in the absence of the vortices, we find that the  vortex-state-mediated critical supercurrent is enhanced with an amount of about $33$ percent in the short junction limit $L=0.1\xi$, which drops rapidly (about 38 percent) as the cylinder length is doubled($L=0.2\xi$) that implies tunneling characteristics between the two vortices. In addition to the pronounced step structures, the supercurrent also declares a reversal as long as  $L\geq\xi$, with a significant suppression of the critical supercurrent when compared to the gedenken junction case.

The number of propagating channels on the cylindrical surface is given by $2k_F R-1$. There exist three nearly degenerate  propagating channels $\nu=0,\pm 1$  in the present case $k_FR=2$, so the critical supercurrent through the gedenken junction can approach  $3e\Delta/\hbar$ in the short junction limit. Significant enhancement of the supercurrent as $k_FL=1$ in the presence of vortex bound states is due to the formation of additional transport channels, which is clearly shown in Fig. \ref{energy-spectra-KR=2}. However, these additional channels with small angular momentum carry a significant supercurrent(they belong to effective transport channels), such as the $|\mu|=\frac52,\frac72,\frac92$ channels,  the others with large angular momentum do not. Formation of these additional transport channels is originated from  effective coherent tunneling between the vortex bound states with small angular momentum.  When the angular momentum of a vortex bound state is coincident with that of an Andreev subband, there would be a hybridization between them, and the three degenerate Andreev subbands split.

The observed step structure can be well understood from the excitation spectra given in Fig. \ref{energy-spectra-KR=2}. With the increase of the superconducting phase difference, more and more effective propagating channels are involved in transporting cooper pairs, resulting in step increase of the supercurrent. If the cylinder length exceeds a critical value, no effective propagating channels can be formed from the vortex state tunneling. However, hybridization between the vortex bound states and the Andreev subband is still possible as long as their angular momenta are coincident, while the other vortex bound states with large angular momentum behave as impurities someway. The hybridization  leads to  splitting of the almost degenerate Andreev subbands  and  modifies the phase difference dependence of the subband energy, especially for the subband with angular momenta $\mu=\pm \frac 32 $.  It is just  such a hybridization that finally leads to a supercurrent reversal, and the impurity-similar effect that leads to the suppression of the critical current\cite{Bagwell}.

\begin{figure}

\centering
		\includegraphics[width=0.4\textwidth]{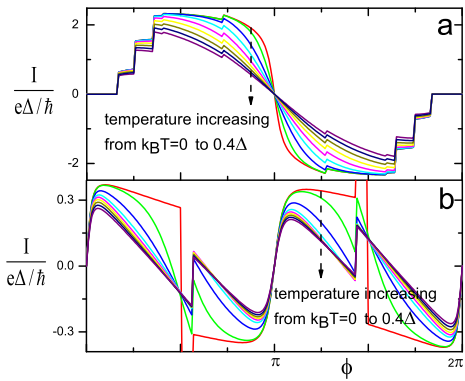}

\caption{Supercurrent-phase relation $I(\phi)$ at different temperatures of the junction with cylinder length (a) $k_FL=2$ and (b) $k_FL=13$. The other parameters are the same  as in Fig. \ref{current-phase-kR=2}.}
\label{temperature-current-phase-kR=2}
\end{figure}

The finite-temperature supercurrent through the junction is given in Fig. \ref{temperature-current-phase-kR=2}, corresponding to Fig. \ref{current-phase-kR=2} at zero temperature in the short junction limit($L=0.2\xi$) and in the intermediate junction case($L=1.3\xi$). At finite temperatures, supercurrent is reduced , while the step structure persists in the short junction limit and develops into a sawtooth oscillation structure in the intermediate junction case. Temperature influences the supercurrent through the combination of Fermi factors $tanh(E/2k_BT)$ in Eq. (\ref{Josephson current}). Notice the fact that the thermally energy-averaging function $tanh(E/2k_BT)$ is more effective with smaller energies, it is not difficult to understand that supercurrent persists its step structure in the short junction limit  and develops into a sawtooth oscillation in the intermediate junction case.

A final remark is on the effect of impurities on the supercurrent-phase evolution. The impurities on the superconductor have already  been  considered by introducing a "dirty-limit" coherence length $\xi_{SC}$. It is known that a single impurity inside the normal part of a one-dimensional SNS junction behaves as a phase modulator and suppresses the supercurrent\cite{Bagwell}. If several impurities are introduced on the normal cylindrical surface, more complicated  supercurrent step structure can be expected. We argue that the supercurrent will be not suppressed significantly by the impurities, since quasiparticles can easily bypass the impurities in our case.

In summary, we have proposed a SNS junction supporting vortex bound states, and found a striking vortex-state-mediated Josephson effect. The vortex-state-mediated supercurrent may evolve in a step or sawtooth fashion  in response to increasing the superconducting phase difference, and its magnitude may be significantly enhanced or suppressed, depending on the junction length. Moreover these striking supercurrent characteristics can not be smeared out by thermal effects and impurity effects, thus provide a smoking-gun evidence for the existence of vortex bound states in type-II superconductors.

Discussions with Yi-bin Huang are gratefully acknowledged.


\section{SUPPLEMENTARY ONLINE MATERIAL}

\begin{widetext}

\section{ Perturbative solutions of the BdG equation}

The Bogoliubov-de-Gennes equation (\ref{BdG-equation}) in the maintext with an inhomogeneous pair potential  $\Delta(\bm{\rho})$ can not be solved exactly. In the extreme Type II limit, the magnetic effect may be neglected.  Equation \label{BdG-equation} is then reduced to

\begin{equation}
\Big\{-\frac{\hbar^2}{2m}
\Big[\frac{1}{\rho}\frac{d}{d\rho}\Big(\rho\frac{d}{d\rho}\Big)-\frac{1}{\rho^2}\Big(\mu^2+\frac14\Big)+k_F^2\Big]\sigma_z
 + \Delta(\rho)\sigma_x  \Big\}
  \Big[\begin{array}{l} u(\rho)\\v(\rho)\end{array}\Big]=\Big(E+\frac{\mu\hbar^2}{2m\rho^2}\Big)\Big[\begin{array}{l} u(\rho)\\v(\rho)\end{array}\Big].
\label{BeG-equation-no-A}
\end{equation}

Consider a radius $r_c$ such that $(\mu+1/2)/k_F\ll r_c\ll\xi$, we seek solutions for $r>r_c$ and separate the two component wavefunction into a rapidly oscillating part $H^{1,2}_{\mu'}(k_F\rho)$ and a slowly varying part $R_{\pm}(k_F\rho)$ with $\mu'=(\mu+1/4)^{1/2}$

\begin{equation}
\Big[\begin{array}{l} u(\rho)\\v(\rho)\end{array}\Big]=\Big[\begin{array}{l} R_+(\rho)\\R_-(\rho)\end{array}\Big]H^{1,2}_{\mu'}(k_F\rho),
\label{fast-slow separation}
\end{equation}
the Hankel functions satisfy the equation $ -H^{''}-H'/\rho+(\mu'^2/\rho^2-k_F^2)H=0$. Since the asymptotic form of the Hankel functions is $H_n^{1,2}(x)=(2/\pi x)^{1/2}\exp[\pm i(x+n^2/2x-n\pi/2-\pi/4)]$ , we have $dH_{\mu'}^{1,2}(k_F\rho)/d\rho
\approx \pm ik_F H_{\mu'}^{1,2}(k_F\rho)$. Substitution of Eq. (\ref{fast-slow separation}) into Eq. (\ref{BeG-equation-no-A}) yields the following equation for the envelop functions

\begin{equation}
\Big[-i\sigma_z \frac{\hbar^2 k_F}{m}\frac {d}{d\rho}+\Delta(\rho)\sigma_x\Big]\Big[\begin{array}{l} R_+(\rho)\\R_-(\rho)\end{array}\Big]
=\Big(E+\frac{\mu\hbar^2}{2m\rho^2}\Big)\Big[\begin{array}{l} R_+(\rho)\\R_-(\rho)\end{array}\Big].
\label{perturbative equation}
\end{equation}

Note that the envelope functions corresponding to the Hankel functions of the first and second kind are complex conjugate each other. For low-energy excitations $E \ll \Delta$, the right-hand side of Eq. (\ref{perturbative equation}) can be treated as a perturbation and it will be solved to first order. The envelope functions can be expressed in the form $[ R_+(\rho),R_-(\rho)]^T=[e^{i\mathcal{S}(\rho)}, e^{-i\mathcal{S}^*(\rho)}]^T$. After substituting it into
Eq. (\ref{perturbative equation}), and expanding $\mathcal{S}(\rho)$ to first order, i.e., $\mathcal{S}=\Re+i\Im=\mathcal{S}_0+\mathcal{S}_1=\Re_0+\Re_1+i(\Im_0+\Im_1)$, we obtain

\begin{eqnarray}
\hbar v_F\frac{d(\Re_0+\Re_1)}{d\rho}+\Delta(\rho)cos[2(\Re_0+\Re_1)]&=&E+\frac{\mu\hbar^2}{2m\rho^2},\nonumber\\
\hbar v_F\frac{d(\Im_0+\Im_1)}{d\rho}-\Delta(\rho)sin[2(\Re_0+\Re_1)]&=&0.
\end{eqnarray}

The zero-order solutions are $\Re_0=\pi/4$, $ \Im_0=\int_0^\rho \Delta(r)dr/\hbar v_F$, and the first-order solutions are
$\Im_1=0$,

\begin{equation}
\Re_1(\rho)=-\int_\rho^\infty e^{2[\Im_0(\rho)-\Im_0(r)]}\Big(\frac {E}{\hbar v_F}+\frac{\mu}{2k_Fr^2}\Big)dr.
\end{equation}

\section{ Derivation of the Hamiltonian for the normal cylindrical surface}

 The Schr\"{o}dinger equation for particles  moving on a flat surface is $i\hbar \partial_t \psi(x,y)=[
  -\hbar^2 \nabla^2/2m+V(x,y)]\psi(x,y)=[-\hbar^2(\partial^2_x+\partial^2_y)/2m+V(x,y)]\psi(x,y)$. If particles are confined ideally to a curved surface embedded in the three-dimensional Euclidean space, the derivative $\partial_j$ must be replaced by its covariant form $D_j$ defined as $D_j A^i=\partial_j A^i+\Gamma^{i}_{jk}A^k$, and the Laplacian $\nabla^2$ by the Laplace-Beltrami operator $g^{-1/2}\partial_i(g^{ij}g^{1/2}\partial_j)$. Here  $g^{ij}=\mathbf{e}^i\cdot\mathbf{e}^j=\nabla u^i\cdot\nabla u^j$ is the metric tensor of the curved space, $\Gamma^{i}_{jk}=\mathbf{e}^i\cdot \partial^2_{jk}\mathbf{r}$ are the Christoffel symbols, and $g=det(g_{ij})$ with $g_{ij}=\mathbf{e}_i\cdot\mathbf{e}_j=\partial_i\mathbf{r}\cdot \partial_j\mathbf{r}$. After parameterizing  the cylindrical surface by $(u^1,u^2)=(\theta,z)$, the position vector reads  $\mathbf{r}=(R cos\theta, R sin\theta,z)$. The components of the metric tensor  are $g^{11}=1/g_{11}=1/\rho^2,g^{22}=g_{22}=1, g^{12}=g^{21}=g_{12}=g_{21}=0$, and all the Christoffel symbols are zero. Then the Schr\"{o}dinger equation for electrons  moving free on the cylindrical surface takes the form  $i\hbar \partial_t \psi(\theta,z)=[-\hbar^2(\partial^2_\theta/R^2+\partial^2_z)/2m]
  \psi(\theta,z)$. It is noted that the above equation differs from that obtained from the thin-layer procedure\cite{Ferrari} by a constant effective potential $-\hbar^2/8mR$, which is trivial and can be absorbed into the Fermi energy.

  \section{ Matching conditions and excitation spectra}

  To determine the excitation spectrum of interest, one faces up with how to  match wavefunctions at a sharp corner. Any real manifold must be smooth and the transition between different charts should be $C^\infty$ differentiable. A proper procedure is to smooth the sharp edge into a quarter-circle of neglected radius. Rotational symmetry of the structure allows us to match the wavefunctions at any  polar angle. To ensure current conservation in the absence of tunneling between the vortex bound states,  wavefunctions and their derivatives  must be continued at the boundaries:$\mp \Psi_C(\theta,z)\partial_z\Psi_C(\theta,z)|_{z=0,L}=\Psi_S(\rho,\theta)\partial_\rho\Psi_S(\rho,\theta)|_{\rho=R}$. The minus sign is due to a reversal of the current flowing on the upper and lower surface.   Taking into consideration the orthogonality relation $\int_{0}^{2\pi}e^{i(n-n')\theta}d\theta=2\pi\delta_{n,n'}$, we obtain a set of equations to match the wavefunctions on the upper(A) and lower(C) superconductive surfaces to the wavefunctions  on the cylindrical surface(B)

\begin{eqnarray}
\label{boundary1}
  A_\mu^1H^{1}_{\mu'}(k_FR)\Big[\begin{array}{l} e^{i\mathcal{S}+i\phi} \\e^{-i\mathcal{S}^*}\end{array}\Big] +A_\mu^2H^{2}_{\mu'}(k_FR)\Big[\begin{array}{l} e^{-i\mathcal{S}^*+i\phi} \\e^{i\mathcal{S}}\end{array}\Big]&=&\left[\begin{array}{l}  B^{1+}_{\nu_1}+B^{1-}_{\nu_1} \\ B^{2+}_{\nu_2}+B^{2-}_{\nu_2} \end{array}\right],\\
\label{boundary2}
A_\mu^1H^{1}_{\mu'}(k_FR)\Big[\begin{array}{l} e^{i\mathcal{S}+i\phi} \\e^{-i\mathcal{S}^*}\end{array}\Big] -A_\mu^2H^{2}_{\mu'}(k_FR)\Big[\begin{array}{l} e^{-i\mathcal{S}^*+i\phi} \\e^{i\mathcal{S}}\end{array}\Big]& =&-\left[\begin{array}{l} \tilde{k}_{+}^{\nu_1}(B^{1+}_{\nu_1}-B^{1-}_{\nu_1}) \\ \tilde{k} _{-}^{\nu_2}(B^{2+}_{\nu_2}-B^{2-}_{\nu_2}) \end{array}\right],\\
\label{boundary3}
    C_\mu^1H^{1}_{\mu'}(k_FR)\Big[\begin{array}{l} e^{i\mathcal{S}} \\e^{-i\mathcal{S}^*}\end{array}\Big] +C_\mu^2H^{2}_{\mu'}(k_FR)\Big[\begin{array}{l} e^{-i\mathcal{S}^*} \\e^{i\mathcal{S}}\end{array}\Big] & =&\left[\begin{array}{l}  B^{1+}_{\nu_1}e^{ik_{+}^{\nu_1}L}+B^{1-}_{\nu_1}e^{-ik_{+}^{\nu_1}L} \\ B^{2+}_{\nu_2}e^{ik_{-}^{\nu_2}L}+B^{2-}_{\nu_2}e^{-ik_{-}^{\nu_2}L} \end{array}\right],\\
\label{boundary4}
    C_\mu^1H^{1}_{\mu'}(k_FR)\Big[\begin{array}{l} e^{i\mathcal{S}} \\e^{-i\mathcal{S}^*}\end{array}\Big] -C_\mu^2H^{2}_{\mu'}(k_FR)\Big[\begin{array}{l} e^{-i\mathcal{S}^*} \\e^{i\mathcal{S}}\end{array}\Big] &=&\left[\begin{array}{l} \tilde{k}_{+}^{\nu_1}(B^{1+}_{\nu_1}e^{ik_{+}^{\nu_1}L}-B^{1-}_{\nu_1}e^{-ik_{+}^{\nu_1}L}) \\ \tilde{k}_{-}^{\nu_2}(B^{2+}_{\nu_2}e^{ik_{-}^{\nu_2}L}-B^{2-}_{\nu_2}e^{-ik_{-}^{\nu_2}L}) \end{array}\right],
  \end{eqnarray}
where $\nu_{1,2}=\mu\mp1/2$, and $\tilde{k}_{\pm}^{\nu_i}=k_{\pm}^{\nu_i}/k_F$ are valued at $\rho=R$. To obtain this set of matching equations, we have also used  the fact that $dH_{\mu'}^{1,2}(k_F\rho)/d\rho
\approx \pm ik_F H_{\mu'}^{1,2}(k_F\rho)$ and $|dS(\rho)/d\rho|\ll k_F$. It seems most convenient to absorb the Hankel functions and the real parts of the exponentials into the corresponding expansion coefficients. By adding and extracting Eqs. (\ref{boundary1}) and (\ref{boundary2}), (\ref{boundary3}) and (\ref{boundary4}),  we obtain two equations for just two expansion coefficients, say $B^{1\pm}$, after some simple algebraic manipulations. The resulting equation to determine the excitation  spectrum is

\begin{eqnarray}
&& \big[1-\tilde{k}_{+}^{\nu_12}\big]
\big[[1-\tilde{k}_{-}^{\nu_22} \big]sin(k_{+}^{\nu_1}L)
sin(k_{-}^{\nu_2}L)+\Big\{4\tilde{k}_{+}^{\nu_1}\tilde{k}_{-}^{\nu_2}cos(k_{+}^{\nu_1}
 L)cos(k_{-}^{\nu_2}L)+\big[1+\tilde{k}_{+}^{\nu_12}\big]\big[[1+\tilde{k}_{-}^{\nu_22}\big]
sin(k_{+}^{\nu_1}L)sin(k_{-}^{\nu_2}L)\Big\}\nonumber \\
&&cos(4\Re_1) +2\Big\{\big[1+ \tilde{k}_{+}^{\nu_12}\big]\tilde{k}_{-}^{\nu_2}sin(k_{+}^{\nu_1}L) cos(k_{-}^{\nu_2}L)-\tilde{k}_{+}^{\nu_1}\big[1+\tilde{k}_{-}^{\nu_22}\big]cos(k_{+}^{\nu_1}L)  sin(k_{-}^{\nu_2}L)\Big\}sin(4\Re_1)+4\tilde{k}_{+}^{\nu_1}\tilde{k}_{-}^{\nu_2}cos\phi=0.\nonumber
\end{eqnarray}
This equation seems  unsatisfactorily complicated. However, in the Andreev approximation $\tilde{k}_{\pm}^{\nu_i}\approx1$, it can be reduced to a desirable form
\begin{equation}
(k_{+}^{\nu_1}-k_{-}^{\nu_2})L-4\Re_1(R)\pm\phi=(2n+1)\pi.
\end{equation}

\section{ Andreev spectra for the Gedenken SNS junction}

In this section, we use the scattering matrix method to derive  the Andreev spectra for a Gedenken SNS junction. The Gedenken junction is an idealization of the structure described in the maintext, allowing a uniform pair potential for $\rho\geq R$  while prohibiting the appearance of vortex bound states. The Bogoliubov-de-Gennes equation with fixed angular momentum $\mu$ becomes

\begin{eqnarray}
\Big\{- \frac{\hbar^2}{2m}
\Big[\frac{1}{\rho}\frac{d}{d\rho}\Big(\rho\frac{d}{d\rho}\Big)-\frac{\mu^2}{\rho^2}+k_F^2\Big]\sigma_z
 + \Delta(cos\phi \sigma_x -sin\phi\sigma_y)\Big\}   \Big[\begin{array}{l} u(\rho)\\v(\rho)\end{array}\Big]=E\Big[\begin{array}{l} u(\rho)\\v(\rho)\end{array}\Big].
\label{BdG-equation-gedenken}
\end{eqnarray}

The eigenfunctions of single electron(hole) Hamiltonian$
\mathscr{H}_{e,h}=\mp\hbar^2[d_\rho(\rho d_\rho)/\rho-
\mu^2/\rho^2]/2m$ are just Hankel functions $\psi^{e,h}_k(\rho)=H^{1,2}_{\mu}(k\rho)$, and satisfy the relation $\mathscr{H}_{e,h}H^{1,2}_{\mu}(k\rho)=\pm \hbar^2k^2 H^{1,2}_{\mu}(k\rho)/2m$. Expressing the two-component wave functions $[u(\rho),v(\rho)]^T$ in the form $[ae^{i\phi},b]^TH^{1,2}_{\mu}(k\rho)$, we obtain
the eigenvalues of the Bogoliubov-de-Gennes equation
\begin{equation}
E=\pm\sqrt{\Big[\frac{\hbar^2(k^2-k_F^2)}{2m}\Big]^2+\Delta^2},
\label{eigen-value-BdG}
\end{equation}
where $\pm$ symbols the electronlike and holelike branch of the excitation spectrum. The constituting electron- and hole- amplitudes for these two-branch wavefunctions are

\begin{eqnarray}
E&>&0: a=\sqrt{\frac{\Delta}{2E}}e^{\frac{1}{2}cos^{-1}\frac{E}{\Delta}},\hspace{0.4cm} b=\sqrt{\frac{\Delta}{2E}}e^{-\frac{1}{2}cos^{-1}\frac{E}{\Delta}};\\
E&<&0: a=\sqrt{\frac{\Delta}{2E}}e^{-\frac{1}{2}cos^{-1}\frac{E}{\Delta}};\hspace{0.2cm}  b=\sqrt{\frac{\Delta}{2E}}e^{\frac{1}{2}cos^{-1}\frac{E}{\Delta}}.
\end{eqnarray}

 The wave vectors of electronlike and holelike excitations are given by $k^{\pm}=k_F\sqrt{1\pm \sqrt{E^2-\Delta^2}/E_F}$. Therefore the scattering state of the superconductors can be expanded as $\psi_e^{S\pm}=[ue^{i\phi},v]H^{1,2}_{\mu}(k_+\rho)$ and  $\psi_h^{S\pm}=[ve^{i\phi},u]H^{1,2}_{\mu}(k_-\rho)$.
For the cylindrical surface Hamiltonian, The propagating electronlike and holelike modes on the normal cylindrical surface are described by $\psi_e^{\pm}(z)=[1,0]^Te^{\pm ik^ez}$ and  $\psi_h^{\pm}(z)=[0,1]^Te^{\pm ik^hz}$, $k^{e,h}=k_F^\mu\sqrt{1\pm E/E_F^\mu}$, where $E_F^\mu=E_F-\hbar^2\mu^2/2mR^2$ and
 $ k_F^\mu=\sqrt{2mE_F^\mu}/\hbar$. In the Andreev approximation,  normal reflection probability is too  small to be considered. An electron-like excitation $\psi_{e}^{S-}$, incidenting at $\rho=R$ from the upper superconductive surface, may be Andreev reflected as a hole-like quasiparticle with wavefunction $r_{eh}\psi_{h}^{S-}$ onto the same surface, or transmitted onto the other lower superconductive surface with $t_{ee}\psi_{e}^{S+}$. The supporting propagation models on the cylindrical surface are $a\psi_e^+(z)+b\psi_h^+(z)$.  Connecting these scattering states at the boundaries yields the transmission amplitude
\begin{equation}
t_{ee}=\frac{(u^2-v^2)H_{\mu}^{2}(k^+R)}{H_{\mu}^{1}(k^+R)(u^2e^{i\phi}e^{-ik^eL}-v^2e^{-ik^hL})}.
\end{equation}
The Andreev spectrum is then determined by the resonance condition
\begin{equation}
-2cos^{-1}\frac{E}{\Delta}\pm\phi+(k^e-k^h)L=2n\pi.
\label{gedenke-Andreev-spectrum}
\end{equation}

\section{ Derivation of the Josephson current formula}

We give an alternative and thorough derivation of the Josephson current formula, starting from a simple  excitation scenario usually adopted in condensed matter physics.

The Josephson current is an equilibrium thermodynamical property of superconductors. It can be associated with the variation of some kind of thermodynamic potential with respect to vector potential or superconducting phase variation. Consider a single-particle Hamiltonian $H=m\mathbf{\hat{v}}^2/2+V(\mathbf{r})=(\mathbf{\hat{p}}-q\mathbf{A}/c)^2/2m+V(\mathbf{r})$ and an infinitesimal variation $\delta A$, the variation of the Hamiltonian is $\delta \hat{H}=-q\mathbf{\hat{v}}\cdot \delta \mathbf{A}/c=-q\int d\mathbf{r} (|\mathbf{r}><\mathbf{r}|\mathbf{\hat{v}}+\mathbf{\hat{v}}|\mathbf{r}><\mathbf{r}|)\cdot \delta \mathbf{A}/2c=-\int d\mathbf{r} \mathbf{\hat{J}}\cdot \delta \mathbf{A}/c$. Then the single-particle current density operator $\mathbf{\hat{J}}$ equals to $-c\delta \hat{H}/\delta \mathbf{A}$. Now we turn to many particle canonical ensemble systems, the quantum ensemble average of the current density is $\mathbf{J}=\langle \mathbf{\hat{J}}\rangle=tr[\mathbf{\hat{J}}e^{-\hat{H}/k_BT}]/tre^{-\hat{H}/k_BT}=c\delta F/\delta \mathbf{A}$, $F=-k_BTlntre^{-\hat{H}/k_BT}$ is the free energy. According to the gauge theory of superconductors\cite{Gennes}, variation of the vector potential $\delta \mathbf{A}$ will induce a variation in the gradient of the superconducting phase $\frac{\hbar c}{2e}\delta \nabla \phi$. The current density may then be written equivalently  as $\mathbf{J}=\frac{2e}{\hbar}\frac{\delta F}{\delta \nabla\phi}$. Since $\delta F=\int d\mathbf{r} \frac{\delta F}{\delta \nabla\phi}\cdot\delta \nabla\phi=\frac{\hbar}{2e}\int  d\mathbf{s}\cdot d\mathbf{l}_\phi \mathbf{J}\cdot \delta \nabla\phi= \frac{\hbar}{2e}I\delta \int d\mathbf{l}\cdot \nabla \phi=\frac{\hbar}{2e}I\delta \int_l d\phi=\frac{\hbar}{2e}I\delta \Delta \phi$,  the Josephson current is therefore written in a familiar form\cite{Bardeen Beenakker}  $I=\frac{2e}{\hbar} \frac{\partial F}{\partial \Delta\phi}$.

The mean-field BCS Hamiltonian for a phase-gradient inhomogeneous superconductor may be generally written as $\hat{H}=E_0+\sum_{q\sigma} E_q(\Delta \phi) :\hat n_{q\sigma}:$, where$: :$represents normal ordering, $E_0$ is the ground state energy independent of the superconducting phase, $E_q$ is the excitation energy and generally depends on the gradient of the superconducting phase. The canonical partition function is $Z=tre^{-\hat{H}/k_BT}=e^{-E_0/k_BT}e^{\sum_qE_q/k_BT}\prod_{q,\sigma}(1+e^{-E_q/k_BT})$, and the free energy yields $F=E_0-2k_BT\sum_q ln[2cosh(E_q/2k_BT)]$. Since we have normally ordered the Hamiltonian, the summation over $q$ should be counted from the ground state energy, i.e., $E_q>0$. One can also alternatively derive the free energy from statistical physics. $F=U-TS=U+k_BT\sum_{i\sigma} [f_{i\sigma}lnf_{i\sigma}+(1-f_{i\sigma})ln(1-f_{i\sigma})]$, where $f_{i\sigma}=1/[exp(E_{i\sigma}/k_BT)+1]$ is the Fermi-Dirac distribution function . Simple algebra yields $F=U+\sum_i (-E_i)f_{i\sigma}+\sum_{i}E_i (1-f_{i\sigma})-2k_BT\sum_i ln[2cosh(E_i/2k_BT)]=E_0-2k_BT\sum_q ln[2cosh(E_q/2k_BT)]$. The above analysis indicates that, the Josephson current flows along the direction of the gradient of the superconducting phase, and can be expressed as

\begin{eqnarray}
I=-\frac{2e}{\hbar}\sum_{q}^{0< E_q<\Delta}tanh\Big(\frac{E_q}{2k_BT}\Big)\frac{\partial E_q}{\partial\Delta\phi}- \frac{2e}{\hbar}\int_{\Delta}^{\infty}dE  \rho_S(E) tanh\Big(\frac{E}{2k_BT}\Big)\frac{\partial E}{\partial\Delta\phi}.
\end{eqnarray}

The first and second terms represent respectively the contributions from the discrete subgap
Andreev levels and the continuum states above the gap.

\end{widetext}

\end{document}